\documentclass[review]{elsarticle}

\usepackage[utf8]{inputenc}
\usepackage{ulem}
\usepackage{color,graphicx}
\usepackage{amssymb,amsmath}
\usepackage{bm,bbm,hyperref}
\usepackage{verbatim}
\usepackage{gensymb}
\usepackage{float}
\usepackage{soul}
\usepackage{physics,natbib}
\usepackage{bbold}
\usepackage{mathrsfs}
\usepackage[varg]{txfonts} 
\usepackage{lineno,hyperref}

\newcommand{\be}{\begin{eqnarray}}
	\newcommand{\ee}{\end{eqnarray}}
\newcommand{\beq}{\begin{equation}}
	\newcommand{\eeq}{\end{equation}}
\newcommand{\eq}[1]{\begin{align}#1\end{align}}

\newcommand{\mc}{\mathcal}
\newcommand{\mbb}{\mathbbm}
\newcommand{\mf}{\mathfrak}
\newcommand{\mr}{\mathrm}

\begin{document}

\begin{frontmatter}

\title{Two-time quantities as elements of physical reality}

\author{Lucas Maquedano$^{1}$, Alexandre D. Ribeiro$^{1,2}$, Ana C. S. Costa$^{1}$, and Renato M. Angelo$^{1}$}
\address{$^{1}$ Department of Physics, Federal University of Paran\'a, 81531-980 Curitiba, PR, Brazil. \\
$^{2}$ Department of Physics, University of Connecticut, Storrs, Connecticut CT 06269, USA.}


\begin{abstract}
In recent years, time correlators have received renewed attention, especially under the guise of identifiers of nonclassical correlations. However, the physical interpretation of these objects, and more generally of multi-times variables, remains ambiguous, which may be one of the reasons why they are so difficult to measure. In this work, we introduce and advance the perspective that a two-time correlator should actually be regarded as an average involving a novel single physical observable, one that cannot be rephrased in terms of the primitive ones, according to quantum principles. In particular, we provide examples showing that the presumed constituents of a two-time correlator and the proposed two-time operator itself cannot be simultaneous elements of the physical reality.
\end{abstract}


\end{frontmatter}



\section{Introduction}

Classical physics relies on a foundational assumption: because (i) all physical quantities have well-defined observer-independent values at a given instant of time (realism) and (ii) physical systems adhere to cause-effect relations (local causality), then physical quantities must have well-defined observer-independent values for every other instant of time, whether in the past or the future (determinism). Quantum mechanics stands in direct opposition to this view; its algebraic structure, leading to the uncertainty principle~\cite{Heisenbeg}, already prohibits simultaneous realism for position and momentum, thus banishing determinism. Moreover, as evidenced by the debate sparked by the seminal articles of Einstein, Podolsky, and Rosen (EPR)~\cite{EPR}, Bell~\cite{Bell64,Brunner2014}, and Bohm~\cite{Bohm1,Bohm2}, which ended up in several conclusive tests~\cite{Hensen15,Shalm15,Giustina15}, nature does not allow for the coexistence of realism and local causality. 

Violations of further classical-like premises have also been observed~(see the reviews \cite{Emary2014,Giuseppe2023}) with respect to the approach proposed by Leggett and Garg~\cite{Leggett-Garg,Emary2014}. These authors pointed out that quantum mechanics is at odds with macrorealism---the assumption that macroscopic physical quantities (e.g., the position of a massive object) have definite values at every instant of time and that such values can be experimentally accessed with arbitrarily small disturbances. The Leggett-Garg inequalities, whose violation has become the standard notion of nonclassical temporal correlations, are based on two-time correlators of the form $\langle\mc{A}_1\mc{B}_2\rangle\equiv \Tr\big( C_{12}^\times\,\rho_0\big)$, where $\rho_0$ is a quantum state (a preparation) and $C_{12}^\times\equiv C^\times(t_1,t_2)\coloneqq \frac{1}{2}\{A_1,B_2\}$. Here, $A_1\equiv A(t_1)$ and $B_2\equiv B(t_2)$ stand for Hermitian Heisenberg operators associated with the respective physical quantities $\mc{A}_1\equiv\mc{A}(t_1)$ and $\mc{B}_2\equiv\mc{B}(t_2)$, defined at times $t_1$ and $t_2$, respectively. Symmetrization (via anticommutator) is welcome because $A_1B_2\neq B_2A_1$ in general, even when $A_0B_0=B_0A_0$. Although this article restricts itself to projective measurements only, it is worth mentioning that the anticommutator does not generally result in a positive operator~\cite{Fritz2010}.

Now, time-correlation functions are ubiquitous in science, being commonly used to characterize the dynamics of stochastic processes. Particularly noteworthy in recent years has been the intense effort to apply the so-called out-of-time-order correlators~\cite{Brian2018,Green2022} as quantitative tools in the study of quantum chaos~\cite{garcia-mata23} and information propagation in quantum many-body systems~\cite{he2017,swingle2017}. However, the interpretation of these objects, and even of simpler versions like $\Tr\big(C_{12}^\times\rho_0\big)$ and $\Tr\big(C_{12}^+\rho_0\big)$, with $C_{12}^+\coloneqq A_1+B_2$, is tricky from a fundamentally quantum viewpoint. First, notice that $\Tr\big(C_{12}^\times\rho_0\big)$, in particular, does not admit a clear statement in the Schr\"odinger picture, implying that it is picture-dependent, in contrast with most mean values typically encountered in physics. Second and more seriously, both formulations seem to impose upon us the idea that their statistics have to be inevitably obtained through noninvasive sequential measurements. However, in the quantum domain, measurements are always invasive to some extent.

In this work, we develop a genuinely quantum formulation for two-time operators, which are functions of Heisenberg operators evaluated in two different instants of time, as $C_{12}^\times$ and $C_{12}^+$, treating them as `nonlocal-in-time physical quantities' possessing subtle elements of reality. Specifically, we employ a criterion of realism and an associated quantifier~\cite{Bilobran15} to argue that the two-time operators should not be interpreted simply as compound quantities but rather as distinct quantum concepts in their own right. The rest of the paper is organized as follows. To motivate and justify our work, in Sect.~\ref{sec2}, we raise some conceptual difficulties to manage time-correlators. Then, we develop our ideas in Sect.~\ref{sec3}, identifying the two-time observables and characterizing them using the concept of realism. At last, we present our final remarks in Sect.~\ref{sec4}.

\section{Two-point measurement criticism}
\label{sec2}

Induced by classical intuition, one might be tempted to interpret the two-time correlator $\langle \mc{A}_1\mc{B}_2\rangle$ as a weighted average of elements of reality $a_1$ and $b_2$, occurring at times $t_1$ and $t_2$, respectively, over some probability distribution. In this perspective, experimental tests would follow from direct measurements of the physical quantities $\mc{A}$ and $\mc{B}$ at the pertinent times. In the quantum paradigm, on the other hand, one can take for granted neither the concept of pre-existing elements of reality nor the concept of noninvasive measurements. In this section, we want to explicitly state the fact that, in general, quantum mechanics forbids one from considering that two-time quantities can be experimentally accessed through a two-point measurement (TPM) protocol.

Let us consider quantities $\mc{A}$ and $\mc{B}$, whose quantum counterparts are nondegenerate discrete-spectrum Hermitian operators given by $A=\sum_a a\,\alpha_a$ and $B=\sum_b b\,\beta_b$, with projectors $\alpha_a$ and $\beta_b$, respectively. In addition, we assume a preparation $\rho_0$ and a time evolution induced by an arbitrary map $\phi_t$ satisfying $\phi_0=\mbb{1}$, $\phi_{t_1}\phi_{t_2}=\phi_{t_1+t_2}$, and, exclusively for unitary operations, $\phi_t^*=\phi_{-t}$. With that, we write $A_1\equiv A(t_1)\equiv \phi_{t_1}(A)$ and $B_2\equiv B(t_2)\equiv \phi_{t_2}(B)$, in Heisenberg's picture, and $\rho_t=\phi_t^*(\rho_0)$, in Schr\"odinger's picture. In terms of the unitary time-evolution operator $U_t$, one has $\phi_t^*(\rho_0)\equiv U_t\rho_0 U_t^\dag$. If $\mc{A}$ is measured at $t_1$, the state $\rho_{t_1}$ jumps to $\alpha_a$ with probability $p(a,t_1) =\Tr\big[\alpha_a\phi_{t_1}^*(\rho_0)\big]$. A sequential measurement of $\mc{B}$, at $t_2>t_1$, yields the state $\beta_b$ with (conditional) probability $p(b,t_2|a,t_1)=\Tr\big[\beta_b\phi_{t_2-t_1}^*(\alpha_a)\big]$. So, according to this TPM protocol, the resulting correlator reads
\eq{\label{A1B2_TPM}
	\langle \mc{A}_1\mc{B}_2\rangle^\text{\tiny TPM} &= \sum_{a,b}ab\,p(b,t_2|a,t_1)p(a,t_1) \nonumber \\
	&=\sum_{a,b}ab\,\Tr\big[\beta_b\phi_{t_2-t_1}^*(\alpha_a)\big]\Tr\big[\alpha_a\phi_{t_1}^*(\rho_0)\big],
}
which requires a time ordering for the measurements, but no symmetrization. On the other hand, for the Heisenberg formulation of the two-time correlator, one considers $A_1=\sum_aa\,\phi_{t_1}(\alpha_a)$ and $B_2=\sum_bb\,\phi_{t_2}(\beta_b)$ to write
\eq{\label{A1B2_Heis}
	\langle \mc{A}_1\mc{B}_2\rangle & =\tfrac{1}{2}\langle \{A_1,B_2\} \rangle\nonumber \\ &=\tfrac{1}{2}\sum_{a,b}ab\,\Tr\big[\{\phi_{t_1}(\alpha_a),\phi_{t_2}(\beta_b)\}\,\rho_0\big] \nonumber \\
	&=\tfrac{1}{2}\sum_{a,b}ab\,\Tr\big[\phi_{t_2}(\beta_b)\,\{\phi_{t_1}(\alpha_a),\rho_0\}\big] \nonumber \\
	&=\tfrac{1}{2}\sum_{a,b}ab\,\Tr\big[\beta_b\,\phi_{t_2}^*(\{\phi_{t_1}(\alpha_a),\rho_0\})\big] \nonumber \\
	&=\sum_{a,b}ab\,\Tr\big[\beta_b\phi_{t_2-t_1}^*(\Lambda_{t_1})\big]\Tr\big[\alpha_a \phi_{t_1}^*(\rho_0)\big],
}
where we have introduced, for the sake of comparison with Eq.~\eqref{A1B2_TPM}, the operator
\eq{\Lambda_{t_1}=\frac{\phi_{t_1}^*(\{\phi_{t_1}(\alpha_a),\rho_0\})}{2\Tr[\phi_{t_1}(\alpha_a)\, \rho_0]}=
	\frac{\{\alpha_a,\rho_{t_1}\}}{2\Tr(\alpha_a \rho_{t_1})}.
}
The differences between this operator and $\alpha_a$ are remarkable. The first one is the explicit dependence on $t_1$ and $\rho_0$. In the TPM protocol, any details about the preparation and the state evolution up to $t_1$ are completely erased by the first measurement, and their effects are exclusively captured by the quantity $\Tr\big[\alpha_a \phi_{t_1}^*(\rho_0)\big]$. Second, and more surprisingly, $\Lambda_{t_1}$ is not even a valid quantum state in general. This implies that, by forcing the two-time correlator to mimic the results of the TPM protocol in this way, one may end up with absurdities. To see this, consider the dynamics of a qubit in the Bloch sphere. If we set $\alpha_a=(\mbb{1}+a\hat{z}\cdot\vec{\sigma})/2$ and $\rho_0=(\mbb{1}+\hat{r}_0\cdot\vec{\sigma})/2$, with $\vec{\sigma}$ the vector composed of the Pauli matrices, $a=-1$, and $||\hat{r}_0||=1$, then we find, for some generic unitary evolution, a resulting pure state $\rho_{t_1}=\phi_{t_1}^*(\rho_0)=(\mbb{1}+\hat{r}_1\cdot\vec{\sigma})/2$, where $\hat{r}_1$ is a versor expressed by some function of $t_1$. Direct calculation leads to $\Lambda_{t_1}=(\mbb{1}+\vec{\nu}_1\cdot\vec{\sigma})/2$, with $\vec{\nu}_1=(\hat{r}_1-\hat{z})/(1-\hat{z}\cdot\hat{r}_1)$. Finally, using $\hat{z}\cdot\hat{r}_1=\cos\theta$, one demonstrates that $||\vec{\nu}_1||=|\sin(\theta/2)\,|^{-1}\geq 1$, with equality holding only for $\theta=\pi$. That is, $\Lambda_{t_1}$ will generally be outside the Bloch sphere---a nonphysical state.

The above calculation does not definitively close the discussion about the nonequivalence between expressions~\eqref{A1B2_TPM} and~\eqref{A1B2_Heis}. However, it is appropriate to identify a peculiar scenario in which the TPM protocol agrees with the Heisenberg formulation for $\langle \mc{A}_1\mc{B}_2\rangle$. This is the case for a time-evolved state such that
\eq{\label{BA_realism}\rho_{t_1}=\Phi_A(\rho_{t_1})\coloneqq\sum_a \alpha_a\,\rho_{t_1} \alpha_a.
}
Here, it readily follows that $[\alpha_a,\rho_{t_1}]=0$, so that $\Lambda_{t_1}=\alpha_a$. What is particularly interesting about Eq.~\eqref{BA_realism} is that, according to a given criterion (to be briefly reviewed in the next section), it means that the quantity $\mc{A}$ is an element of reality at~$t_1$. Conceptually, this explains why both procedures successfully converge in this case: the first measurement does not determine reality by force but rather probes a pre-existing one. The question then arises as to whether there can be some element of reality associated with a two-time observable $\frac{1}{2}\{A_1,B_2\}$, a concept that does not involve any intermediary measurement. In the next section, we put the focus on measurements aside to develop a theory addressing this issue.

\section{Two-time elements of reality} 
\label{sec3}

Recently, one of us and a coauthor advanced the perspective according to which mechanical work is an observable delocalized in time~\cite{Silva21,Silva23}, that is, a Hermitian operator in Heisenberg's picture equipped with an eigensystem defined for fixed times $t_{1,2}$. Inspired by these ideas, here we propose to take the two-time Heisenberg operator $C_{12}$, assumed as $C_{12}^\times$ or $C_{12}^+$, as a new observable {\it per se}. In other words, $C_{12}$ is assumed to be the quantum mechanical representation of a corresponding physical quantity that possesses elements of reality in no way constrained by the elements of reality of $\mc{A}_1$ and $\mc{B}_2$. 

To start with, let us borrow the EPR notion of an element of reality~\cite{EPR}, understood as a definite value for a given physical quantity. With this preliminary concept in mind, we study a simple instance of $C_{12}^+$. For a particle of mass $m$ moving freely along the $x$ axis, during the time interval $\Delta t = t_2-t_1$, the displacement operator reads
\eq{\delta_{12}\coloneqq X_2-X_1=P\,\Delta t/m,
}
where $X_k=\phi_{t_k}(X)=X+Pt_k/m$, with $k\in\{1,2\}$, and $[X,P]=i\hbar\mbb{1}$. From the well-known inequality $\Delta A\,\Delta B\geq \frac{1}{2}|\langle [A,B]\rangle|$ and the relations $[X_1,X_2]=[X_k,\delta_{12}]=i\hbar\mbb{1} \Delta t/m$, one has $\Delta X_1\,\Delta X_2\geq \hbar \Delta t/(2m)$ and
\eq{\Delta\delta_{12}\left(\Delta X_1+\Delta X_2 \right)\geq \frac{\hbar\Delta t}{m}.
}
This result shows that there is no preparation able to make $\delta_{12}$, $X_1$, and $X_2$ simultaneous elements of reality. In particular, it accepts scenarios in which $\Delta\delta_{12}\to 0$ and $\Delta X_{1,2}\to \infty$. In words, the displacement can be an element of reality while the constituent positions are not. This materializes, for instance, when the particle is prepared in a Gaussian state with a generic mean momentum $p_0$ and negligible momentum uncertainty $\Delta p$. It follows from $\Delta \delta_{12}=\Delta p\,\Delta t/m$ that for any bounded time interval, displacement will be definite.

This simple example provides us with at least two lessons. First, quantum mechanics allows for scenarios in which the degree of definiteness of $\delta_{12}$ is not constrained by the degrees of definiteness of its defining positions. Hence, $\delta_{12}$ can be prepared to be an independent element of reality. Second, the word ``prepared'' is important here because the two-time operator $\delta_{12}$ has no physical meaning at any specific instant of time. This requires extending the notion of elements of reality to a ``nonlocal-in-time'' perspective: by preparing a momentum eigenstate $\ket{p_0}$ at $t=0$, we ascertain that the element of reality $p_0(t_2-t_1)/m$ will manifest for the particle's displacement after the time interval $[t_1,t_2]$ has elapsed. Notice that such a notion of a nonlocal-in-time element of reality is required for displacement also within the classical paradigm, the difference being that in classical physics, the positions are elements of reality at all times. In addition, while in classical physics the displacement can be certified by sequential position measurements at $t_1$ and $t_2$, in quantum mechanics the invasiveness of measurements forbids the TPM protocol to provide certification for displacement, as discussed previously. In this example, the first position measurement would collapse the quantum state to a position eigenstate, making the displacement $P\Delta t/m$ maximally indefinite and, therefore, not an element of reality.

To further emphasize the first point discussed above, let us consider the dynamics induced by the Hamiltonian $H=\omega S_z$ on a spin-1/2, where $\omega$ is Larmor frequency. It is an exercise to show that $S_x(t_1)=\frac{\hbar}{2}\hat{u}_1\cdot\vec{\sigma}$ and $S_y(t_2)=\frac{\hbar}{2}\hat{v}_2\cdot\vec{\sigma}$, with $\hat{u}_1=(\cos\tau_1,-\sin\tau_1,0)$, $\hat{v}_2=(\sin\tau_2,\cos\tau_2,0)$, and $\tau_k=\omega t_k$. The only possible elements of reality for $S_x(t_1)$ and $S_y(t_2)$ are $\pm \frac{\hbar}{2}$, for all $t_{1,2}$. Now, if the elements of reality of $\frac{1}{2}\{S_x(t_1),S_y(t_2)\}$ were to emerge from simple products of the elements of reality of $S_x(t_1)$ and $S_y(t_2)$, we would expect only $\pm\frac{\hbar^2}{4}$ as outcomes. However, $\frac{1}{2}\{S_x(t_1),S_y(t_2)\}=\frac{\hbar^2}{4}\hat{u}_1\cdot\hat{v}_2\mbb{1}\!=\!0$, a Hermitian operator with null eigenvalues whose only element of reality expected is zero.

Of course, what we learned so far is not specific to the chosen observables. Formally,  Heisenberg operators can always be expanded in terms of a summation of Schr\"odinger operators, with each parcel of the sum being multiplied by some function of the fixed times $t_{1,2}$. Then, any well-behaved two-time operator $C_{12}$ can be diagonalized and have its eigenvalues $c^{12}_k$ and eigenvectors $|c^{12}_k\rangle$ found. Since in Heisenberg's picture the quantum state does not evolve in time, upon preparing $\rho_0=|c^{12}_k\rangle\langle c^{12}_k|$, we ensure that the element of reality $c^{12}_k$ will have manifested after the time lapse $[t_1,t_2]$.

\subsection{Quantifying violations of realism}

To emphasize our argument that $C_{12}$ can be viewed as a two-time element of reality, we now consider an improved criterion of realism. EPR associate elements of reality with situations in which one knows the value of a physical quantity for certain even before making any measurement of it~\cite{EPR}. Basically, this amounts to having a quantum state $\alpha_a=|a\rangle \langle a|$ corresponding to an eigenstate $\ket{a}$ of an observable $A$ representing a physical quantity $\mc{A}$. More recently, a criterion was proposed~\cite{Bilobran15} that allows one to diagnose realism even in scenarios involving an observer's subjective ignorance. Suppose that $A=\sum_aa\alpha_a$ is measured on a system prepared in the state~$\rho$. The post-measurement state reads  $\alpha_a\rho \alpha_a/\Tr(\alpha_a\rho)=\alpha_a$. Hereafter, $A$ is an EPR element of reality, since we can predict with certainty the outcome of any further measurement of $A$ even without measuring it effectively. Now, let us assume that, for some reason, the observer forgets the measurement outcome $a$. Because this process occurred in the observer's brain, the system's ontology should not change. Therefore, the remaining description $\sum_a p_a \alpha_a$, with $p_a=\text{Tr}(\alpha_a\rho)$, simply reflects an epistemic ignorance about a pre-established ontology. In other words, $\sum_ap_a \alpha_a=\sum_a\alpha_a\rho \alpha_a=\Phi_A(\rho)$ is a state for which $A$ is an element of reality. In short, $\Phi_A(\rho)$ is called an $A$-reality state. With that, the authors of Ref.~\cite{Bilobran15} propose the equality $\rho=\Phi_A(\rho)$ [see Eq.~\eqref{BA_realism}] as a criterion of $A$-realism. The rationale is direct: if a preparation $\rho$ is not altered by a nonselective measurement of $A$ (a measurement followed by forgetting), then the measurement is innocuous in establishing realism; an element of reality associated with $A$ had already been installed in the preparation. The authors then moved one step further and proposed the so-called $A$ {\it irreality},
\eq{\label{mfI} \mf{I}(A|\rho)\coloneqq \mr{S}\big(\Phi_A(\rho)\big)-\mr{S}(\rho)\equiv \min_\sigma\mr{S}\big(\rho||\Phi_A(\sigma)\big),
}
where $\mr{S}(\rho||\eta)\coloneqq\Tr[\rho(\ln\rho-\ln\eta)]$ is the relative entropy of $\rho$ and $\eta$, and $\mr{S}(\rho)\coloneqq-\Tr(\rho\ln\rho)$ is the von Neumann entropy of $\rho$. Since $\mf{I}(A|\rho)\geq 0$, with equality holding iff $\rho=\Phi_A(\rho)$, it is clear that the $A$ irreality measures how far the observable $A$ is from being an element of reality for the preparation $\rho$. As a simple example, consider a two-level system with a preparation $\Omega=\ket{0}\!\bra{0}$. In this case, while $S_z$ is an element of reality, as $\mf{I}(S_z|\Omega)=0$, $S_x$ is maximally unreal, since $\mf{I}(S_x|\Omega)=\ln{2}$. It is clear that the quantum coherence of the state $\Omega$ in the $S_x$ basis is the mechanism that forbids this observable to be an element of reality. As shown in Ref.~\cite{Bilobran15}, when $\rho$ is multipartite, not only coherence contributes to the violation of $A$ realism, but also correlations do. It is fair to say that irreality is by now a well-established concept, supported by numerous investigations, both theoretical~\cite{Dieguez2018,Gomes2018,Gomes2019,Fucci2019,Orthey2019,Engelbert2020,Costa2020,Savi2021,Orthey2022,Paiva2023,Engelbert2023} and experimental~\cite{Mancino2018,Dieguez2022}.

Referring back to our previous discussion, consider the preparation   $\rho_0=\sum_kp_k|c^{12}_k\rangle\langle c^{12}_k|$, expanded in the eigenbasis of $C_{12}$ with a probability distribution $p_k$. Direct calculations show that $\rho_0=\Phi_{C_{12}}(\rho_0)$, implying $\mf{I}(C_{12}|\rho_0)=0$. In words, a preparation such as $\rho_0$ guarantees, according to the irreality measure, that $C_{12}$ will be an element of reality, while $A_1$ and $B_2$ do not need to be. In the next section, we give a concrete illustration of the general discussion conducted so far.

\subsection{Application: Torque on a spin-1/2}

Let us consider the unitary dynamics of a spin-1/2 in a constant magnetic field aligned with a generic direction $\hat{h}$. The coupling is modeled as $H = \omega S_h$, with $S_h \equiv \frac{\hbar}{2}\,\hat{h} \cdot \vec{\sigma}$. By using textbook results regarding the Pauli matrices, the Heisenberg time evolution $\vec{\sigma}(t)=U_t^\dag\vec{\sigma}U_t$ of $\vec{\sigma}$ is found to be
\eq{\label{sigma_t}\vec{\sigma}(t)= \vec{\sigma}\cos\tau  + \hat{h} \times \vec{\sigma}\sin\tau  + \hat{h}\,\hat{h} \cdot \vec{\sigma}(1-\cos\tau)\equiv \vec{\sigma}_\tau,
}
where $\tau=\omega t$. We now introduce the dimensionless torque imposed on the spin:
\eq{\vec{T}(t_1,t_2)\coloneqq \frac{2}{\hbar \omega}\frac{\vec{S}(t_2)-\vec{S}(t_1)}{t_2-t_1}=\frac{\vec{\sigma}(t_2)-\vec{\sigma}(t_1)}{\tau_2-\tau_1}.
}
Our aim here is to show that $\vec{T}(t_1,t_2)$ and $\vec{\sigma}(t_{1,2})$ can never be simultaneous elements of reality, except in a fairly classical regime. To this end, it will be enough to set $\tau_1=\omega t$ and $\tau_2=\omega (t+\Delta t)$, with $\Delta t\to 0$. In this limit, $\vec{T}(t_1,t_2)$ reduces to the instantaneous torque
\eq{
	\vec{T}_\tau =\hat{h}\times \vec{\sigma}\,\cos{\tau} +(\hat{h}\,\hat{h}\cdot \vec{\sigma}-\vec{\sigma})\,\sin{\tau}.
}
By $\hat{h}\cdot\vec{T}_\tau=0$, we see that $\vec{T}_\tau$ is orthogonal to $\hat{h}$, as physically expected. For simplicity, we set $\tau=2\pi$ and $\hat{h}=\hat{z}$, so that the objects of interest reduce to $\vec{T}_{2\pi}=\hat{z}\times\vec{\sigma}$ and $\vec{\sigma}_{2\pi}=\vec{\sigma}$. Our last specialization consists of focusing on the $x$ component of the torque and the corresponding component of the dimensionless angular momentum:
\eq{\label{x-Tsigma} T_{2\pi}^x\equiv \hat{x}\cdot\vec{T}_{2\pi}=\sigma_y,\qquad \quad\sigma_{2\pi}^x\equiv \hat{x}\cdot \vec{\sigma}_{2\pi}=\sigma_x.
} 

\begin{figure}
	\centerline{\includegraphics[scale=0.6]{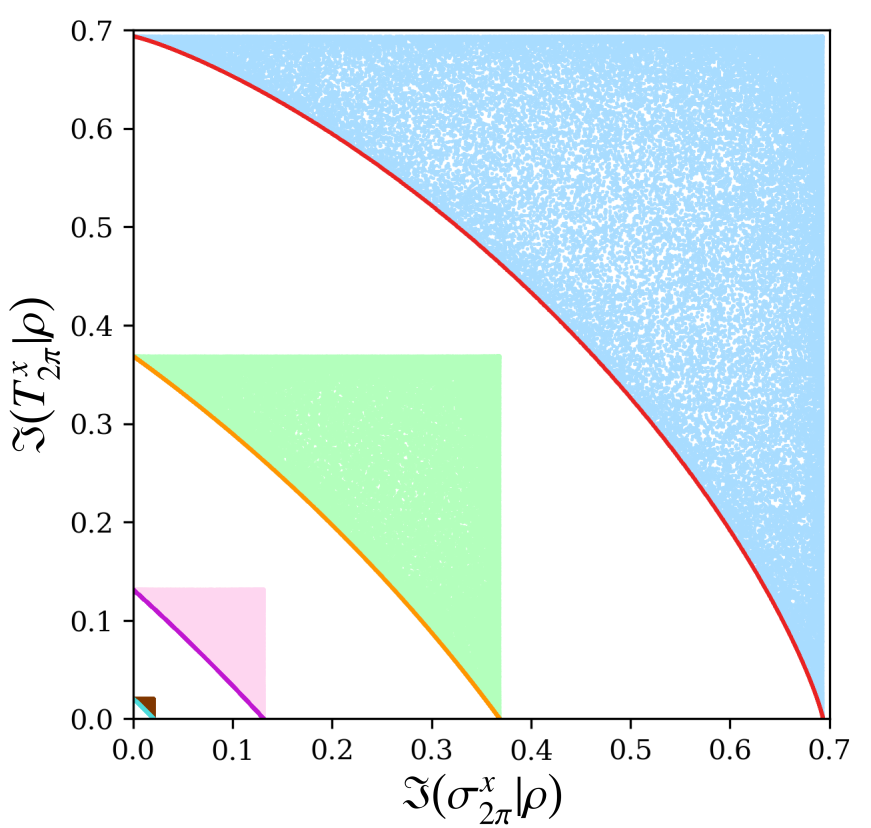}}
	\caption{Parametric plot of $\mf{I}(T_{2\pi}^x|\rho_0)$, the irreality of the $x$ component of the torque, versus $\mf{I}(\sigma_{2\pi}^x|\rho_0)$, the analogous for the spin angular momentum, for states $\rho_0 = \frac{1}{2}(\mathbbm{1} + \vec{r}\cdot \vec{\sigma})$ parametrized by the vector $\vec{r} = r\{\sin{\theta}\cos{\phi},\sin{\theta}\sin{\phi},\cos{\theta}\}$, with $\theta\in[0,\pi]$, $\phi\in[0,2\pi)$, and $r\leq 1$. Each of the four bands is composed of points corresponding to the aforementioned irrealities computed for $10^4$ randomly generated $(\theta,\phi)$ pairs. Each band is associated with different values of $r$. From the bottom-left to top-right corners, we have $r=0.2$ (brown points forming the band near the origin), $r=0.5$ (pink points forming the band above the first), $r=0.8$ (green points), and $r=1.0$ (blue points). The solid curves, which bound each band from below, refer to states lying on the equator of the Bloch sphere, corresponding to $\theta=\pi/2$.}
	\label{fig1}
\end{figure}

Intuitively, we can anticipate that being noncommuting operators, the two-time observable $T_{2\pi}^x$ and its constituent $\sigma_{2\pi}^x$ cannot be simultaneous elements of reality\footnote{It is noteworthy that, according to the EPR's approach, there exist scenarios (involving entangled states) in which $\sigma_x$ and $\sigma_y$ can be simultaneous elements of reality.}. To confirm this, we now compare the irrealities $\mf{I}\big(T_{2\pi}^x|\rho_0\big)$ and $\mf{I}\big(\sigma_{2\pi}^x|\rho_0\big)$ for a generic quantum state $\rho_0 = \frac{1}{2}(\mbb{1} + \vec{r}\cdot \vec{\sigma})$, parameterized by $\vec{r} = r\{\sin{\theta}\cos{\phi},\sin{\theta}\sin{\phi},\cos{\theta}\}$, with $\theta\in[0,\pi]$, $\phi\in[0,2\pi)$, and $r\leq 1$. The analytical results read
\begin{subequations}
	\eq{\mf{I}(T_{2\pi}^x|\rho_0)&=\mr{H}_\text{bin}\left(\tfrac{1+|\hat{y}\cdot\vec{r}|}{2}\right)-\mr{H}_\text{bin}\left(\tfrac{1+r}{2}\right), \\
		\mf{I}(\sigma_{2\pi}^x|\rho_0)&=\mr{H}_\text{bin}\left(\tfrac{1+|\hat{x}\cdot\vec{r}|}{2}\right)-\mr{H}_\text{bin}\left(\tfrac{1+r}{2}\right),
	}
\end{subequations}
where $\mr{H}_\text{bin}(\mf{u})\coloneqq-\mf{u}\ln{\mf{u}}-(1-\mf{u})\ln{(1-\mf{u})}$ stands for the binary entropy, with $\mf{u}\in[0,1]$. Figure \ref{fig1} shows a parametric plot of the aforementioned irrealities for $10^4$ randomly generated states, for each of the four values of $r$: $r=0.2$ [brown points forming the band close to the origin], $r=0.5$ (pink points forming the band above the first), $r=0.8$ (green points forming the band above of the previous), and $r=1.0$ (blue points forming the band above the last). Note that the parameter $r$ is a direct measure of the state purity, given by $\Tr\rho_0^2=\frac{1}{2}(1+r^2)$. The solid curves, bounding each band from below, refer to states lying on the equator of the Bloch sphere, those for which $\theta=\pi/2$. Since the eigenstates of the operators under scrutiny also lie on the equator, preparations with $\theta=\pi/2$ naturally minimize the sum of irrealities. The numerical results clearly show that, except for $r=0$ (the maximally mixed state\footnote{Because $\mf{I}(O|\mbb{1}/d)=0$ for any Hilbert space dimension $d$, the maximally mixed state is said ``classical'', as it allows all observables $O$ to be elements of reality simultaneously.}), the irrealities can never be simultaneously zero, meaning that $T_{2\pi}^x$ and $\sigma_{2\pi}^x$ can never be simultaneous elements of reality.

An analytical proof of this last numerical result can be provided using a distinctive property of relative entropy~\cite{Umegaki1962}, known as the data-processing inequality, $\mr{S}\big(\Lambda(\rho)||\Lambda(\varrho)\big)\leq \mr{S}(\rho||\varrho)$, which is also called contractivity or monotonicity under quantum
channels $\Lambda$ (completely positive trace-preserving maps). Setting $\varrho=\Phi_{\sigma_x}(\rho)$ and $\Lambda=\Phi_{\sigma_y}$, so that $\Lambda(\varrho)=\Phi_{\sigma_y}\circ\Phi_{\sigma_x}(\rho)=\frac{\mbb{1}}{2}$, we find
\eq{
	\mr{S}\big(\rho||\Phi_{\sigma_x}(\rho)\big) &\geq \mr{S}\big(\Phi_{\sigma_y}(\rho)||\Phi_{\sigma_y}\circ\Phi_{\sigma_x}(\rho) \big) \nonumber \\
	\mr{S}\big(\Phi_{\sigma_x}(\rho)\big)-\mr{S}(\rho)&\geq \mr{S}\big(\Phi_{\sigma_x}\circ\Phi_{\sigma_y}(\rho)\big)-\mr{S}\big(\Phi_{\sigma_y}(\rho)\big) \nonumber \\
	\mr{S}\big(\Phi_{\sigma_x}(\rho)\big)+\mr{S}\big(\Phi_{\sigma_y}(\rho)\big)&\geq \ln{2}+\mr{S}(\rho).
}
Then, with the result $\mr{S}(\rho_0)=\mr{H}_\text{bin}\big[(1+r)/2\big]$, and relations \eqref{mfI} and \eqref{x-Tsigma}, we finally obtain
\eq{\mf{I}(T_{2\pi}^x|\rho_0) + \mf{I}(\sigma_{2\pi}^x|\rho_0) \geq \ln{2}-\mr{H}_\text{bin}\left(\tfrac{1+r}{2}\right).
}
The lower bound in this inequality---a non-negative monotonically increasing function of the state purity---is tight and vanishes only if $r=0$. It formally makes the intended point, namely, that $T_{2\pi}^x$ and $\sigma_{2\pi}^x$ can never be simultaneous elements of reality, except in classical regime ($r=0$).

\section{Conclusions} 
\label{sec4}

Typical formulations of quantum postulates assert that physical quantities are represented by observables (Hermitian operators whose eigenbases span the state space) and that the possible measurement outcomes are the eigenvalues and eigenvectors of these observables. Do these prescriptions apply to all physically meaningful quantities, even those that are not defined at a single instant of time? What, after all, are the eigensystems of multi-time quantities? Given that typical measurements are fundamentally invasive and localized in time, how can one measure a quantity that is delocalized in time? Is it possible to reformulate the Heisenberg picture of a multi-time observable into Schr\"odinger's picture, which would possibly involve a multi-time quantum state? These are some of the intricate questions that arise when we attempt to apply standard quantum mechanics to describe multi-time quantities.

In this work, we discuss how to address the issue within the context of two-time operators, specifically in their simplest forms as $A_1+B_2$ and $\frac{1}{2}\{A_1,B_2\}$ (two-time correlators). Most importantly, we argue that two-time elements of reality can be prepared (instead of measured) and that they cannot be viewed as emerging from the elements of reality of their constituent quantities. Besides complementing the standard quantum formalism, we hope that our results may help researchers advance tests of time correlators.

\section{Acknowledgments}
This study was financed in part by the Coordena\c{c}\~ao de Aperfei\c{c}oamento de Pessoal de N\'ivel Superior - Brasil (CAPES) - Finance Code 001. R.M.A. and A.D.R. thank the financial support from the National Institute for Science and Technology of Quantum Information (CNPq, INCT-IQ 465469/2014-0). R.M.A and A.C.S.C thank the Brazilian funding agency CNPq under Grants No. 305957/2023-6 and 308730/2023-2, respectively.


\bibliography{references}

\end{document}